# Implications of Artificial Intelligence on Health Data Privacy and Confidentiality

Ahmad K. Momani

Department of Computer Science, University of Wisconsin-Milwaukee, Milwaukee WI 53211 USA

*Abstract*

*The rapid integration of artificial intelligence (AI) in healthcare is revolutionizing medical diagnostics, personalized medicine, and operational efficiency. However, alongside these advancements, significant challenges arise concerning patient data privacy, ethical considerations, and regulatory compliance. This paper examines the dual impact of AI on healthcare, highlighting its transformative potential and the critical need for safeguarding sensitive health information. It explores the role of the Health Insurance Portability and Accountability Act (HIPAA) as a regulatory framework for ensuring data privacy and security, emphasizing the importance of robust safeguards and ethical standards in AI-driven healthcare. Through case studies, including AI applications in diabetic retinopathy, oncology, and the controversies surrounding data sharing, this study underscores the ethical and legal complexities of AI implementation. A balanced approach that fosters innovation while maintaining patient trust and privacy is imperative. The findings emphasize the importance of continuous education, transparency, and adherence to regulatory frameworks to harness AI's full potential responsibly and ethically in healthcare.*

**Introduction**

In the digital transformation era, the use of artificial intelligence (AI) in healthcare has surged, from clinical applications in areas such as imaging and diagnostics to operational optimization in hospitals to using health software to assess a patient's symptoms (1). This rapid growth in AI adoption is not just a trend but a promising revolution in healthcare. Many experts believe that AI will not only improve patient care and diagnosing efficiency but also pave the way for a more personalized and effective healthcare system. The economic forecast for the AI healthcare market in the coming years is nothing short of explosive, reflecting the high expectations and optimism surrounding this technology (2).

AI includes fields such as machine learning, deep learning, natural language processing, and robotics. These fields can be applied to any medical field, and their contribution to medical research and healthcare delivery are limitless (3). AI's robust ability to leverage large amounts of clinical data has revolutionized medical diagnostics, decision-making, and personalized medicine. However, at the same time, it also increases ethical concerns about patient confidentiality and privacy (4).

However, this integration comes with many risks and challenges, and AI must be integrated with the healthcare system ethically and legally, particularly concerning the privacy and confidentiality of health data (5,6). The privacy of patient information is a top priority and is guaranteed under laws and regulations such as the Health Insurance Portability and Accountability Act (HIPAA) in the United States (7). HIPAA is a crucial regulatory framework that ensures the security and privacy of health information, protecting patient information against fraud and abuse. As AI technologies can access and analyze vast amounts of data, the importance of HIPAA in safeguarding patient information cannot be overstated (8).

The robustness of AI systems, combined with their ability to learn from clinical data, necessitates the implementation of sturdy safeguards and ongoing monitoring to prevent unauthorized access to data and guarantee adherence to ethical standards regarding data privacy (9). This paper explores the importance of data privacy and confidentiality in healthcare, considering the increasing role of technology and maintaining the balance between harnessing AI's capabilities in healthcare. In addition, this paper addresses the technological, ethical, and legal frameworks that significantly impact the current state of AI in healthcare.

**HIPAA Overview**

The Health Insurance Portability and Accountability Act (HIPAA) is a regularity overseen by the US Department of Health and Human Services (HHS) and is designed to protect patient health information from being used or disclosed without the patient's consent. HIPAA is a crucial regulatory framework in the United States, and its primary goal is to ensure the security and privacy of health information and protect patient information against fraud and abuse (10–13).

HIPAA establishes several mandatory standards for protecting health information. These standards are categorized into critical areas covering various aspects of data management. The Privacy Rule is one of the most essential standards; it sets standards for protecting identifiable health information (14). It defines the situations in which covered entities may use or disclose an individual's protected health information (PHI) (10,14,15). Complementing this, the Security Rule defines the administrative, technical, and physical safeguards that covered entities must employ to ensure the confidentiality of health information (16).

Moreover, the Breach Notification Rule mandates that covered entities notify parties after a breach of PHI. This rule sets the timing and the process of notifications that must be sent to individuals and HHS (17). Lastly, the Enforcement Rule involves comprehensive rules that form the backbone of HIPAA's framework, ensuring that patient data is handled with the highest care and confidentiality (18).

HIPAA significantly affects healthcare providers' daily operations; they must implement policies and procedures to handle PHI securely. Healthcare organizations must train their employees on HIPAA regulations to ensure policy implementation and continuous compliance (11). In addition, they must ensure that any new technology used to store or transmit PHI complies with HIPAA security provisions. This requirement usually requires a significant investment in secure technology systems (19).

**The Role of Technology in Healthcare**

The advancements in technology in the healthcare industry have witnessed a significant transformation in patient care delivery and management. From the development of electronic health records (EHR) to the integration of AI and telemedicine, all steps have been aimed at enhancing diagnostic efficiency, accuracy, and ultimately, patient outcomes.

Technology integration in healthcare began with the development of the HER systems in the late 20th century (20). EHR systems were designed to store, manage, and transfer patient information digitally. These systems significantly improved the accessibility of patient information and communication between different healthcare providers (21). In addition, developing sophisticated diagnostic tools, such as medical imaging technologies and laboratory information systems, enhances patient health outcomes (22). For example, MRI machines give physicians deeper insights into patient conditions, allowing for more accurate diagnosis and timely intervention (23).

More recently, AI has begun to play a crucial role in healthcare (24). AI algorithms are used to analyze complex health data and can assist in diagnosing and predicting diseases, forecasting outcomes, and personalizing medicine (2,3,25). AI-driven tools are increasingly used to predict and identify at-risk patients and provide preventive recommendations. This proactive approach also reduces healthcare costs by preventing severe health incidents (5,26,27). However, AI integration in the healthcare industry presents significant challenges, particularly regarding patient data privacy and security (28). The increasing use of EHRs and reliance on AI-powered tools expose patient data to potential cyber threats and data breaches (29). Compliance with HIPAA is crucial to safeguard patient data against unauthorized access or misuse.

Healthcare providers and technology experts are constantly working to improve and implement advanced encryption technologies and secure data transmission protocols to boost data security (30). In addition, with the continued progress and development of technology and the increased use of AI in medical research, healthcare providers and researchers should continually educate and train on data privacy practices and the importance of protecting patient information (31).

**AI and Machin Learning in Healthcare**

Machine learning (ML), a branch of AI, is becoming increasingly important in the healthcare industry. It revolutionizes how healthcare professionals diagnose, treat, and conduct clinical research. These algorithms not only improve diagnostic efficiency and health service effectiveness but also bring forth new capabilities for personalized medicine (32).

AI algorithms excel at pattern recognition, which is crucial for interpreting complex medical images such as X-rays, CT scans, and MRIs. AI-driven diagnostic tools can identify subtle patterns that may be invisible to the naked eye (33). For instance, AI algorithms have been developed to detect and analyze abnormalities in radiological images more accurately and quickly than conventional methods, thereby aiding in the early diagnosis of diseases (34). In treatment, AI and ML algorithms personalize medical care based on historical data. AI-powered decision-support systems can suggest customized treatment plans based on the patient's variations in genetics, environment, and lifestyle (35).

AI has significantly benefited clinical research. It can analyze large datasets and uncover trends and associations that might not be apparent through traditional data analysis tools (36). AI models are instrumental in genetic research and help decode genetic markers linked to specific diseases and treatment responses (37). In addition, the capability of the AI model is crucial in epidemiological studies and clinical trials, where researchers need to analyze a vast amount of data to draw meaningful conclusions (38).

The widespread use of AI tools for patient data analysis has raised significant privacy concerns. To address these concerns, it is vital to establish strict data protection measures and ensure that all AI applications comply with relevant health data privacy regulations, such as HIPAA.

**Legal Implications of AI in Healthcare**

Integrating AI into the healthcare industry poses a multitude of regulatory, privacy, and ethical challenges that require thorough examination. As AI continues to transform the healthcare sector, comprehending the existing regulatory framework and ethical ramifications of AI applications is essential to upholding patient confidence and adhering to legal requirements.

HIPAA is the foremost regulation in the United States that ensures the privacy and security of patient information. AI-based tools employed in healthcare must comply with HIPAA's directives and standards, which clearly outline how patient data is managed, shared, and protected. Notably, the Privacy Rule and Security Rule of HIPAA play a crucial role in safeguarding patients' information by requiring appropriate measures to be taken to secure their data and providing a roadmap for responding to data breaches. Countries worldwide have their own privacy regulations that impact the use of AI in healthcare. For example, Canada's PIPEDA and the UK's Data Protection Act impose requirements that affect the use of AI with healthcare data (39,40).

It is essential to provide patients with sufficient information regarding the use of their data. Informed consent is a significant ethical consideration, especially when artificial intelligence algorithms are employed to process their information for diagnosis and treatment purposes. It is vital to be transparent about the functionalities and limitations of AI systems to ensure patient confidence and trust (41).

Determining liability in AI-driven healthcare is complex. When an AI system makes a decision that results in a medical error, it is challenging to pinpoint accountability. This uncertainty can impact patient rights and trust (42). Therefore, conducting regular privacy impact assessments and audits can help identify and mitigate risks associated with AI systems (43). These practices are essential for complying with regulatory requirements and building and maintaining trust with patients and the public.

Developers of artificial intelligence tools in the healthcare industry should prioritize incorporating privacy and data protection principles during the design stage. This approach ensures that patient privacy is given due consideration and is not treated as an afterthought but integrated into the architecture of AI systems (44). Furthermore, healthcare professionals and AI developers should possess in-depth knowledge of ethical considerations and regulatory requirements associated with AI. Providing continuous education and training can help ensure that all stakeholders understand their responsibilities and the ethical implications of AI use (45).

**Cases and Examples**

AI has significantly improved patient outcomes and driven research breakthroughs within the healthcare industry. However, its use has raised several ethical and legal concerns. This section includes case studies highlighting the benefits and challenges associated with AI in healthcare.

*1. Early Detection of Diabetic Retinopathy.* AI has made significant strides in healthcare, particularly in the diagnosing of diabetic retinopathy, a severe complication that can cause blindness in diabetic patients (46). Google Health has created an advanced AI system that can analyze eye scans and accurately detect signs of diabetic eye disease with a level of accuracy comparable to human ophthalmologists (47). This breakthrough technology has the potential to detect and treat the condition earlier, thereby saving the vision of countless patients worldwide. This case study is a prime example of how AI can improve diagnostic accuracy and ultimately improve patient outcomes.

*2. AI in Oncology Research.* IBM Watson for Oncology was created to aid cancer treatment by analyzing structured and unstructured data in clinical notes and reports using context and meaning (48). The system delivers personalized treatment recommendations by comparing various treatment options against medical literature, clinical practices, and patient information. This AI application has been crucial in advancing oncology research, providing insights that refine treatment protocols and improve patient survival rates. However, despite its promise to revolutionize cancer treatment, IBM Watson has faced significant challenges and criticism, particularly around its effectiveness and the quality of its recommendations. Reports suggest that the system sometimes provides unsafe and incorrect cancer treatment advice. This situation underscores the potential risks of relying on AI for crucial health decisions without adequate oversight and validation. The ethical implications concerning patient safety and the accuracy of AI-driven decisions have come to the forefront, raising concerns about the deployment of AI in life-or-death scenarios.

*3. **DeepMind Health Data Sharing Controversy***. The DeepMind Health Data Sharing Controversy is another example of AI and healthcare facing public scrutiny. DeepMind, a subsidiary of Alphabet, came under fire for its partnership with the UK's National Health Service, which involved sharing patient data without explicit consent to develop an app to detect kidney injury (49). The lack of transparency and failure to sufficiently inform patients about how their data was being used led to a significant backlash and raised questions about privacy, consent, and trust in using AI in healthcare.

These case studies highlight the benefits and challenges of AI in healthcare. AI can improve diagnostic accuracy, treatment options, and medical research. Nevertheless, issues about data privacy, the accuracy of AI recommendations, and the lack of transparency can compromise trust and jeopardize patient well-being.

Healthcare providers and AI developers must comply with strict ethical standards and regulatory requirements to overcome these challenges. This involves thorough testing, ensuring transparency in AI operations, and obtaining informed consent for data usage. Additionally, it is essential to continuously monitor and evaluate the performance of AI applications in healthcare to ensure they remain safe, effective, and in the best interest of patients.

**Conclusion**

The integration of AI in healthcare has the potential to revolutionize patient care and drive significant breakthroughs in medical research. The case studies mentioned earlier have demonstrated that AI can effectively improve diagnostic accuracy, personalize treatment options, and detect diseases at an early stage. However, these advancements also introduce new ethical and data management complexities, emphasizing the critical need for robust regulatory frameworks and operational transparency.

To effectively address these challenges, a balanced approach is required to promote innovation while upholding patient privacy and trust. Ethical concerns, particularly those related to data privacy, bias, and accountability in AI decision-making, must be rigorously managed through strict adherence to regulations such as HIPAA. Moreover, healthcare providers and AI developers must engage in ongoing dialogue with regulatory bodies, patients, and the public to ensure that AI tools are used responsibly and ethically.

**References**

1. Smith M. Ethical Application of AI [Internet]. Taxodiary. [cited 2024 May 1]. Available from: https://taxodiary.com/2021/11/ethical-application-of-ai/
2. Artificial Intelligence in Healthcare | Accenture [Internet]. [cited 2024 Apr 20]. Available from: https://www.accenture.com/au-en/insights/health/artificial-intelligence-healthcare
3. Ramesh AN, Kambhampati C, Monson JRT, Drew PJ. Artificial intelligence in medicine. Ann R Coll Surg Engl. 2004 Sep;86(5):334.
4. Hagendorff T. The Ethics of AI Ethics: An Evaluation of Guidelines. Minds Mach. 2020 Mar 1;30(1):99–120.
5. Jiang L, Wu Z, Xu X, Zhan Y, Jin X, Wang L, et al. Opportunities and challenges of artificial intelligence in the medical field: current application, emerging problems, and problem-solving strategies. J Int Med Res. 2021;49(3):1–11.
6. Curzon J, Ann Kosa T, Akalu R, El-Khatib K. Privacy and Artificial Intelligence. IEEE Trans Artif Intell. 2021 Apr 1;2(2):96–108.
7. Applied Sciences | Free Full-Text | Balancing Privacy and Progress: A Review of Privacy Challenges, Systemic Oversight, and Patient Perceptions in AI-Driven Healthcare [Internet]. [cited 2024 May 1]. Available from: https://www.mdpi.com/2076-3417/14/2/675
8. Summary of the HIPAA Security Rule | HHS.gov [Internet]. [cited 2024 Apr 21]. Available from: https://www.hhs.gov/hipaa/for-professionals/security/laws-regulations/index.html

9.  Mennella C, Maniscalco U, De Pietro G, Esposito M. Ethical and regulatory challenges of AI technologies in healthcare: A narrative review. Heliyon. 2024 Feb 2;10(4):26297.
10. Privacy | HHS.gov [Internet]. [cited 2024 Apr 30]. Available from: https://www.hhs.gov/hipaa/for-professionals/privacy/index.html
11. HIPAA for Professionals | HHS.gov [Internet]. [cited 2024 Apr 30]. Available from: https://www.hhs.gov/hipaa/for-professionals/index.html
12. Guidance On HIPAA » Fides Tech Solutions [Internet]. 2022 [cited 2024 May 8]. Available from: https://fidestechsolutions.com/services/compliance/guidance-on-hipaa/
13. HIPAA Archives - [Internet]. 2021 [cited 2024 May 2]. Available from: https://osha-safety-training.net/tag/hipaa/
14. HIPAA Privacy Rule and Its Impacts on Research [Internet]. [cited 2024 May 1]. Available from: https://privacyruleandresearch.nih.gov/pr_06.asp
15. HHS Guidance Clarifies Post-Dobbs HIPAA Privacy Protections [Internet]. [cited 2024 May 8]. Available from: https://www.shrm.org/topics-tools/news/benefits-compensation/hhs-guidance-clarifies-post-dobbs-hipaa-privacy-protections
16. Onc H. Guide to Privacy and Security of Electronic Health Information. 2015.
17. American Medical Association [Internet]. 2024 [cited 2024 May 2]. HIPAA Breach Notification Rule. Available from: https://www.ama-assn.org/practice-management/hipaa/hipaa-breach-notification-rule
18. HIPAA violations & enforcement | American Medical Association [Internet]. [cited 2024 May 1]. Available from: https://www.ama-assn.org/practice-management/hipaa/hipaa-violations-enforcement
19. Division HIP. Individuals' Right under HIPAA to Access their Health Information 45 CFR § 164.524 [Internet]. 2016 [cited 2024 May 8]. Available from: https://www.hhs.gov/hipaa/for-professionals/privacy/guidance/access/index.html
20. Tang PC, McDonald CJ. Electronic Health Record Systems. 2006;447–75.
21. Bowman S. Impact of Electronic Health Record Systems on Information Integrity: Quality and Safety Implications. Perspect Health Inf Manag [Internet]. 2013 [cited 2024 May 1];10(Fall). Available from: /pmc/articles/PMC3797550/
22. Hardy M, Harvey H. Artificial intelligence in diagnostic imaging: Impact on the radiography profession. Br J Radiol [Internet]. 2020 Apr 1 [cited 2024 May 1];93(1108). Available from: https://www.birpublications.org/doi/10.1259/bjr.20190840
23. Abuzaid MM, Tekin HO, Reza M, Elhag IR, Elshami W. Assessment of MRI technologists in acceptance and willingness to integrate artificial intelligence into practice. Radiography. 2021 Oct 1;27:S83–7.
24. Koval V. DevSkiller - TalentTech solution for staffing, talent management and engineering teams. 2023 [cited 2024 April 20]. How AI in talent management is revolutionizing HR. Available from: https://devskiller.com/blog/ai-in-talent-management/
25. Stanfill MH, Marc DT. Health Information Management: Implications of Artificial Intelligence on Healthcare Data and Information Management. Yearb Med Inform. 2019 Aug 1;28(1):56–64.
26. Mesko B. The role of artificial intelligence in precision medicine. Expert Rev Precis Med Drug Dev. 2017 Sep 3;2(5):239–41.
27. Ahuja AS. The impact of artificial intelligence in medicine on the future role of the physician. PeerJ. 2019 Oct 4;7:e7702.

28. Williamson SM, Prybutok V. Balancing Privacy and Progress: A Review of Privacy Challenges, Systemic Oversight, and Patient Perceptions in AI-Driven Healthcare. Appl Sci. 2024 Jan;14(2):675.
29. Alanazi A. Clinicians' Views on Using Artificial Intelligence in Healthcare: Opportunities, Challenges, and Beyond. Cureus. 15(9):e45255.
30. Rights (OCR) O for C. Cyber Security Guidance Material [Internet]. 2017 [cited 2024 May 5]. Available from: https://www.hhs.gov/hipaa/for-professionals/security/guidance/cybersecurity/index.html
31. Alowais SA, Alghamdi SS, Alsuhebany N, Alqahtani T, Alshaya AI, Almohareb SN, et al. Revolutionizing healthcare: the role of artificial intelligence in clinical practice. BMC Med Educ. 2023 Sep 22;23(1):689.
32. Alami H, Lehoux P, Auclair Y, de Guise M, Gagnon MP, Shaw J, et al. Artificial Intelligence and Health Technology Assessment: Anticipating a New Level of Complexity. J Med Internet Res. 2020 Jul 7;22(7):e17707.
33. Pesapane F, Codari M, Sardanelli F. Artificial intelligence in medical imaging: threat or opportunity? Radiologists again at the forefront of innovation in medicine. Eur Radiol Exp. 2018 Oct 24;2(1):35.
34. Pathak K, Saikia R, Das A, Das D, Islam MA, Pramanik P, et al. 3D printing in biomedicine: advancing personalized care through additive manufacturing. Explor Med. 2023 Dec 29;4(6):1135–67.
35. Poalelungi DG, Musat CL, Fulga A, Neagu M, Neagu AI, Piraianu AI, et al. Advancing Patient Care: How Artificial Intelligence Is Transforming Healthcare. J Pers Med. 2023 Jul 31;13(8):1214.
36. The Impact of AI in Clinical Data Management - TrialKey [Internet]. 2024 [cited 2024 May 8]. Available from: https://trialkey.ai/blog/the-impact-of-ai-in-clinical-data-management/
37. Abdallah S, Sharifa M, I.KH. Almadhoun MK, Khawar MM, Shaikh U, Balabel KM, et al. The Impact of Artificial Intelligence on Optimizing Diagnosis and Treatment Plans for Rare Genetic Disorders. Cureus. 15(10):e46860.
38. Olawade DB, Wada OJ, David-Olawade AC, Kunonga E, Abaire O, Ling J. Using artificial intelligence to improve public health: a narrative review. Front Public Health. 2023 Oct 26;11:1196397.
39. Sarabdeen J, Chikhaoui E, Mohamed Ishak MM. Creating standards for Canadian health data protection during health emergency – An analysis of privacy regulations and laws. Heliyon. 2022 May 21;8(5):e09458.
40. Canada O of the PC of. Policy Proposals for PIPEDA Reform to Address Artificial Intelligence Report [Internet]. 2020 [cited 2024 May 8]. Available from: https://www.priv.gc.ca/en/about-the-opc/what-we-do/consultations/completed-consultations/consultation-ai/pol-ai_202011/
41. Building Trust in Mental Health AI: The Importance of Transparency | LinkedIn [Internet]. [cited 2024 May 8]. Available from: https://www.linkedin.com/pulse/building-trust-mental-health-ai-importance-scott-1qjuc/
42. Bottomley D, Thaldar D. Liability for harm caused by AI in healthcare: an overview of the core legal concepts. Front Pharmacol. 2023 Dec 14;14:1297353.
43. LLC BTP at TCLBT is P of TC, Advisory A, archiving consulting firm H has 25 plus years in the, Governance I, Privacy D, Security D, et al. Smarsh. 2024 [cited 2024 May 8]. Managing AI to Ensure Compliance with Data Privacy Laws. Available from:

https://www.smarsh.com/blog/thought-leadership/managing-ai-to-ensure-compliance-with-data-privacy-laws
44. Murdoch B. Privacy and artificial intelligence: challenges for protecting health information in a new era. BMC Med Ethics. 2021 Dec 1;22(1).
45. Conrad R. RTS Labs. 2023 [cited 2024 May 8]. Ensuring Ethical Use of AI: Principles, Best Practices, and Implications. Available from: https://rtslabs.com/ensuring-ethical-use-ai-principles-best-practices-implications/
46. Junaid SB, Imam AA, Balogun AO, De Silva LC, Surakat YA, Kumar G, et al. Recent Advancements in Emerging Technologies for Healthcare Management Systems: A Survey. Healthcare. 2022 Oct 3;10(10):1940.
47. Gulshan V, Peng L, Coram M, Stumpe MC, Wu D, Narayanaswamy A, et al. Development and Validation of a Deep Learning Algorithm for Detection of Diabetic Retinopathy in Retinal Fundus Photographs. JAMA. 2016 Dec 13;316(22):2402–10.
48. Strickland E. IBM Watson, heal thyself: How IBM overpromised and underdelivered on AI health care. IEEE Spectr. 2019 Apr;56(4):24–31.
49. Powles J, Hodson H. Google DeepMind and healthcare in an age of algorithms. Health Technol. 2017;7(4):351–67.